\documentclass[aps,10pt,showpacs,twocolumn]{revtex4}
\usepackage{amsmath, amssymb, lipsum}
\usepackage{graphics, graphicx, epsfig, color, subfigure}
\begin{document}

\title{Baryons and Low-Density Baryonic Matter in 1+1 Dimensional Large $N_{c}$  QCD with Heavy Quarks}

\date{\today}

\author{Prabal Adhikari}
\email{prabal@umd.edu}
\affiliation{Maryland Center for Fundamental Physics and the Department of Physics, 
University of Maryland, College Park, MD 20742-4111}

\author{Thomas D. Cohen}
\email{cohen@umd.edu}
\affiliation{Maryland Center for Fundamental Physics and the Department of Physics, 
University of Maryland, College Park, MD 20742-4111}

\author{Arec Jamgochian}
\email{ajamgoch@umd.edu}
\affiliation{University of Maryland, College Park, MD 20742-4111}

\author{Nilay Kumar}
\email{nk2485@columbia.edu}
\affiliation{Columbia University, New York City, NY 10027}
\begin{abstract}
This paper studies baryons and baryonic matter in the combined large $N_{c}$ and heavy quark mass limits of QCD in 1+1 dimension.  This regime allows for  a simple physical picture, which is computationally tractable.  In this non-relativistic limit, baryons are composed of $N_{c}$ quarks that interact, at leading order in $N_{c}$, through a color Coulomb potential. Using variational techniques, single baryon masses and interaction energies of low-density baryon crystal are calculated.  The single baryon calculations are very accurate and can be used as a cross-check to the general numerical approach based on Hamiltonian methods for baryons and baryonic matter in 1+1 dimension for arbitrary quark mass recently  proposed by Bringoltz, which is based on a lattice in a finite box.  It is noteworthy that the Bringoltz method differs from a previous approach developed by  Salcedo, \textit{et al.}  in its treatment of an effect due to the finite box size with periodic boundary conditions -- namely, the existence of gauge configurations which wind around the box.  As these are finite volume effects, one might expect them to be small for boxes that are large enough so that the baryon density approaches zero to high accuracy at the edges of the box.  However, the effects of these windings appear to be quite large even in such boxes.  The large mass infinite volume calculations performed here are consistent with the results of numerical calculations using the Bringoltz approach and confirm the importance of gauge fields that wrap around for finite systems.  The calculation of the baryon crystal interaction energy requires the assumption that at low-densities the ground state is composed of individual baryons, each in color-singlet states, with wave functions orthogonal to one another. While plausible, this assumption is somewhat \textit{ad hoc} in that one can construct configurations in which the entire state is color-singlet but cannot be broken into individual color-singlet baryons. However, the interaction energy of low-density baryon crystals calculated with this assumption is consistent with numerical results based on Bringoltz's approach. This suggests that the assumptions regarding the color structure is justified in 1+1 dimension. This is useful, as it supports a similar assumption that was made in 3+1 dimensions, where there is no alternative means of calculation.
\end{abstract}
\maketitle
\section{Introduction}
The problem of understanding cold nuclear matter from QCD first principles has remained quite challenging due to the intrinsic non-perturbative nature of the problem \cite{borici,politzer}. While many non-perturbative problems such as the finite temperature problem can be studied using lattice QCD techniques, the problem of finite density QCD is marred by the presence of the fermion sign problem \cite{politzer,hands,barbour,alford}; only low densities can be studied and that too only at large temperatures \cite{stepanov} as all available methods are only valid for small values of $\frac{\mu}{T}$. As such it is useful to find certain limiting cases for the study of cold nuclear matter, in which the problem of finite density QCD becomes somewhat more tractable.

Large $N_{c}$ QCD is a limit where the physics simplifies considerably \cite{tHooft1}. Only planar diagrams contribute at leading order in $N_{c}$ while quark loops effects are suppressed. Despite these simplifications, the problem of large $N_{c}$ QCD has remained unsolved. However, the problem in 1+1 dimension is more tractable \cite{tHooft,bars}; here the physics simplifies considerably because gluons, due to the absence of magnetic fields in one spatial dimension, do not propagate. Additionally, with appropriate choice of gauge (either axial or Coulomb), gluon-gluon interactions are absent and hence the theory looks very much like QED, in which color charges (i.e. quarks) interact via a Coulomb potential. The self-energy corrections to the quark propagator come from an infinite number of ladder diagrams, which remarkably can be summed up exactly in spite of the need to solve a bootstrap equation. Furthermore, the meson spectrum can be solved -- albeit only numerically via a Bethe-Salpeter equation; a linear ``Regge trajectory" (with no continuum) is observed.

Unlike the meson problem in 1+1 large $N_{c}$ QCD, the baryon problem is highly non-trivial. Many attempts have been made to study the problem: Ref. \cite{thies} studies the problem in the chiral limit, where baryons assume the form of a Sine-Gordon soliton. Ref.~\cite{narayanan} invokes Eguchi-Kawai reduction~\cite{EK} to argue controversially~\cite{barak} that 1+1 large $N_{c}$ QCD is independent of baryon chemical potential. Ref.~\cite{negele} formulates the large $N_c$ baryon problem for arbitrary quark masses  at infinite volume.  Numerical solutions of the problem require a lattice treatment in a finite-sized box.  The most exhaustive work was done recently by Bringoltz in Ref.~\cite{barak}.  It is based on the axial gauge representation of QCD~\cite{axialQCD} and the coherent states formalism~\cite{yaffe}. Ref.~\cite{barak} solves both the single baryon and baryon crystal problems in a finite sized box for arbitrary quark masses. The principal difference between the approaches of  refs.~\cite{barak} and~\cite{negele} is the treatment of the effects of finite box sizes. In  Ref.~\cite{negele} the infinite volume Hamiltonian is derived and simply used on the finite box by imposing boundary conditions.  In contrast, the  idea underlying Ref.~\cite{barak} is that the quarks making up a baryon in a finite box interact via a color-Coulomb potential of the form $|x-y+QL|$, where $|x-y|$ is the physical separation of the color charges, Q is an integer and L is the size of the box. These additional windings are due to the inability to gauge away the gluon degrees of freedom in the problem~\cite{barak, axialQCD}.  In effect, the treatment of  Ref.~\cite{negele} corresponds to including only the $Q=0$ sector. 

The study of these windings in Ref.~\cite{barak} was motivated largely for theoretical reasons associated with QCD volume reduction.  Clearly in small boxes the two treatments might be expected to be quite different.  {\it A priori}, however, one might expect that the difference between calculations based on Refs.~\cite{barak} and~\cite{negele} for a baryon mass would be quite small in cases where the box sizes are large enough to comfortably contain a baryon since the difference between the two is a finite volume effect.  However, as will be discussed in section \ref{windingsection}, the differences can be quite large, even in such circumstances. Indeed, the differences are large enough that one might worry the differences suggest a possible problem with the formulation of Ref.~\cite{barak}. Unfortunately, It is difficult to verify numerically that this is the case by going to exceptionally large box sizes.

Given this situation, it is quite useful to have a regime in which one can independently compute the mass of a baryon to verify that the winding effects are as large as they appear to be for practical-sized boxes.  The regime of heavy quark masses and large $N_c$ serves this purpose extremely well.  In this paper,  we perform a highly accurate variational calculation of single baryons using Witten's mean-field argument~\cite{witten}, which is similar to an analogous 3+1 dimensional calculation that was done recently in Ref.~\cite{3+1}. It was argued in Ref.~\cite{witten} that baryons are composed of $N_{c}$ quarks in a color singlet state, where each quark moves in an average potential due to the remaining $N_{c}-1$ quarks. In the non-relativistic limit with quark masses taken to infinity, the average potential has a simple form because the only contributions to the interaction come from single gluon exchange.  Note that the principal difference between the heavy quark system using Witten's approach and the general result of Ref.~\cite{barak} is that the vacuum is trivial in the former case.

We find that the single baryon results using Witten's variational formulation agree with the results based on Ref.~\cite{barak} that incle the nontrivial effects of the winding with remarkable accuracy.  These results support the approach taken in Ref.~ \cite{barak} that include a large effect of the windings, albeit only in the restricted regime of large quark masses.  Our result is also quite useful in providing a simple and physically intuitive picture for at least one regime of QCD in 1+1 dimension and, as such, nicely complements the Bringoltz method, which obtains results largely through numerical means.

The study of the heavy quark regime in 1+1 dimension is useful for another purpose: the study of baryonic matter at low density.  Studies of low-density baryonic matter in the heavy quark regime of large $N_{c}$ QCD were previously done in 3+1 dimensions~\cite{3+1}.  The principal attraction of this regime is that it appears to be tractable due to Witten's mean-field arguments and the non-relativistic nature of heavy quark dynamics.  However, as discussed in Ref.~\cite{3+1} for the case of baryonic matter (as opposed to a single baryon) a plausible but {\it ad hoc} assumption was needed: that at low densities the ground state is composed of individual baryons each in a color-singlet state with the wave function of quarks for each baryon orthogonal to the wave function of quarks in any other baryon.  This structure appears quite plausible, but it is  {\it ad hoc} in that one can construct mean-field configurations in which the entire state is color-singlet but cannot be broken into individual color singlet baryons.   In 3+1 dimensions, it was not easy to verify whether this assumption was justified.  However, in 1+1 dimension, one can use the methods of Ref.~\cite{barak} to numerically compute the interaction energy and compare it with the result in the mean-field approach using the assumption of no hidden-color states.  We find good agreement between them, which suggests that this assumption is indeed valid.  This  agreement in 1+1 dimension greatly increases our confidence that the assumption made for the 3+1 dimensional case is valid.
     
The paper is organized as follows. In the following section, we calculate the single baryon energy and wave function in the limit of large quark masses using Witten's mean-field argument. We compare this variational solution to the numerical results based on the recent Hamiltonian approach proposed of Ref.~\cite{barak} and with a previous Hamiltonian in Ref.~\cite{negele} that did not include the windings.  In the next section, we calculate the interaction energy per unit cell of a low density baryon crystal, again using Witten's mean-field argument. We compare the interaction energies with numerical results for the many baryon sector based on Ref.~\cite{barak}.  The paper concludes with a brief discussion of the results.

\section{Baryons in the Combined Large $N_{c}$ and Large $m_{q}$ Limits\label{vari}}
\subsection{The Energy Functional in the Mean Field Approximation}

The nonrelativistic mean-field energy functional for baryons in 3+1 dimensional large $N_{c}$ QCD for heavy quarks  has been calculated in Ref.~\cite{3+1} based on Witten's classic analysis in Ref.~\cite{witten}.   QCD in 1+1 dimension at large $N_{c}$ is generally simpler than in 3+1 dimensions. In 1+1 dimension, quarks are spinless, nonrelativistic fermions and gluon-gluon couplings can be eliminated by appropriate choice of gauge (e.g. the axial gauge as in refs.~\cite{tHooft} and~\cite{negele}). Thus the leading order contribution to the energy comes from a coherent one-gluon exchange between pairs of $N_{c}$ quarks. Interactions involving multiple gluons are suppressed at least at a relative order of $\frac{1}{N_c}$. Furthermore, quarks loops are suppressed at relative order of at least $\frac{1}{N_c}$ as a result of the large $N_{c}$  limit.   However, for the case of heavy quarks, these simplifications do not help.  For heavy quarks, the interactions are pairwise color-Coulomb in any event so  the lack of gluon-gluon interactions adds no new simplifications.  Similarly, spin effects are suppressed for heavy quarks.  Thus, the leading-order energy functional in the dual expansion,  is derived in an essentially identical manner to the 3+1 dimensional case~\cite{3+1}. However, the color-Coulomb potential in the context of one spatial dimension is linear in distance as opposed to dropping off as $\frac{1}{r}$. The strength of the coupling is chosen to be $\frac{g^{2}}{4}$, adopting the conventions in refs.~\cite{tHooft},~\cite{barak} and~\cite{negele}.

The energy of the baryon per color at leading order in $\frac{1}{N_{c}}$ and $\frac{\sqrt{\lambda}}{m_q}$ is 
\begin{align}
&\frac{ \langle\psi|H|\psi\rangle} {N_c}=\frac{M_{\rm baryon}}{N_c}=\\
&m_{q}+\int dx \frac{|\partial_{x}\psi(x)|^2}{2 m_{q}}+\frac{\lambda}{4} \int dx dx' \frac{|\psi(x)|^2 |\psi(x')|^2}{2}|x-x'| \nonumber
\end{align}
where $m_{q}$ is the quark mass, $\psi(x)$ is the normalized baryon wave function and $\lambda\equiv g^{2} N_{c}$.
Minimizing the energy functional with respect to the wave function while imposing the normalization leads to a Schr\H{o}dinger equation for the ground state baryon.
\begin{equation}\begin{split}
\label{se} 
&\left( -\frac{\partial_{x}^{2}}{2 m_{q}}+\frac{\lambda}{4}  \int dx'|\psi_{\rm min}(x')|^2|x-x'|\right)\psi_{\rm min}(x) \\
&=\epsilon\, \psi_{min}(x)\ ,
\end{split}\end{equation}
where $\epsilon$ is the Lagrange multiplier that ensures the normalization of the wave function.

\subsection{The Variational Calculation}
We choose to calculate the ground state baryon wave function $\psi_{\rm min}(x)$ using a variational approach. From here on, $\psi(x)$ will be taken to be the ground state wave function $\psi_{\rm min}(x)$.
To proceed, we define a scaling variable $R$ and a scaled wave function $f(y)$ where $y$ is the dimensionless variable defined as $y\equiv\frac{x}{R}$. Then, the baryon wave function is:
\begin{equation}
\label{three}
\psi(x) = \frac{1}{\sqrt{R}}f\left(y\right)\ .
\end{equation}
The normalization of the wave function $\psi(x)$ is independent of the scaling variable $R$ as long as $f(y)$ is normalized, i.e. $\int dy |f(y)|^2=1$.

Next we define functionals for the kinetic and potential energies, $\mathcal{T}$ and $\mathcal{V}$:

\begin{equation}\begin{split}
&\mathcal{T} [f(y)] \equiv \int dy \frac{|\partial_{y} f(y)|^2}{2}\\
&\mathcal{V} [f(y)] \equiv \int dy dy' \frac{|f(y)|^2|f(y')|^2}{2}|y-y'|\ .
\end{split}\end{equation}

Rewriting the energy functional in terms of the new variables $R$, $\mathcal{T}[f]$ and $\mathcal{V}[f]$, we get:
\begin{equation}\begin{split}
\langle\psi|H|\psi\rangle&=M_{baryon}\\
&=N_{c} \left (m_{q}+\frac{\mathcal{T}[f(y)]}{R^2 m_{q}}+\frac{\lambda}{4} R \mathcal{V}[f(y)]\right)\ .
\end{split}\end{equation}

For simplicity, we choose a normalized functional form for $f(y)$ to be a polynomial times a Gaussian:
\begin{equation}
\label{six}
f(y)= \frac{\left (1+ \sum\limits_{i=1}^{n}a_{i} y^{2i}\right ) e^{\frac{-y^2}{2}}}{\sqrt{\int dy \left (1+\sum\limits_{i=1}^{n}a_{i} y^{2i}\right )^2 e^{-y^2}}}
\end{equation}
where $n$ indicates the number of terms kept. Only even powers of $y$ appear in the wave function because the ground state is expected to be symmetric. We found that truncation at $n=9$ was sufficient to obtain the energy  with a very high accuracy.
We now have an energy functional that is a function of $R$ and $a_{i}$. The next step is to minimize the functional with respect to each of the variables, i.e. $\partial_{h}\langle\psi|H|\psi\rangle=0$ for $h\,\epsilon\,\{R, a_{i}\}$.
Minimizing with respect to $R$ gives
\begin{equation}
\label{R}
R=\left(\frac{8\mathcal{T}[f(y)]}{\lambda m_{q}\mathcal{V}[f(y)]}\right)^{\frac{1}{3}}\ ,
\end{equation}
which yields:
\begin{equation}\begin{split}
\langle\psi|H|\psi\rangle
=N_{c}\left(m_{q}+\frac{3}{4}\mathcal{T}^{\frac{1}{3}}[f(y)]\mathcal{V}^{\frac{2}{3}}[f(y)]m_{q}^{-\frac{1}{3}} \lambda^{2/3}\right)\ .
\end{split}\end{equation}

To proceed we minimize $\langle\psi|H|\psi\rangle$ with respect to each $a_{i}$ for $i = 1 \textrm{ to } 9$ with $a_{0}$ chosen to be $1$. Including $R$, there are a total of 10 parameters in the variational calculation.

The binding energy of the baryon is the second term in Eq. (\ref{energy}) and is characterized by the quantity $\mathcal{T}^\frac{1}{3}\mathcal{V}^\frac{2}{3}$.  It converges rapidly with increasing $n$, the number of parameters, $a_{i}$, in the trial wave function. Minimizing this quantity yields
\begin{equation}\begin{split}
&a_1\approx-0.0450660,\ a_2\approx0.0202162,\ \\&a_3\approx -0.0021246,\ 
a_4\approx0.0001965,\  \\&a_5\approx-6.6218907\times10^{-6},\ a_6\approx-4.4275328\times10^{-8},\ \\&a_7\approx7.8179284\times10^{-9},\ a_8\approx4.2375112\times10^{-11},\ \\&a_9\approx-4.6069215\times10^{-12}\, ,
\end{split} \end{equation}
which fixes the wave function.

It is sensible to present the results in units of $\lambda$ i.e. $g^{2}N_{c}$, which is the only scale present in large $N_{c}$ QCD. Hence masses will be written in the units of $\sqrt{\lambda}$ and lengths in the units of $\frac{1}{\sqrt{\lambda}}$.

The ground state energy per color in units of $\sqrt{\lambda}$ is:
\begin{equation}
\label{energy}
\frac{\langle\psi|H|\psi\rangle}{N_{c}\sqrt{\lambda}}\approx\frac{m_{q}}{\sqrt{\lambda}}+0.25574 \left ( \frac{m_{q}}{\sqrt{\lambda}} \right ) ^{-\frac{1}{3}}\, .
\end{equation}
The variational calculation converged well and the interaction energy is estimated to be accurate to five significant figures.
The parameter $R$ is given by:
\begin{equation}
\label{eleven}
R=\left(\frac{8\mathcal{T}[f(y)]}{\lambda m_{q}\mathcal{V}[f(y)]}\right)^{\frac{1}{3}} \approx 1.079746 \left ( \frac{\lambda}{4} m_{q} \right )^{-\frac{1}{3}}\ ,
\end{equation}
the dimensionless Lagrange multiplier is 
\begin{equation}
\bar{\epsilon}\equiv\left (\frac{\lambda}{4} \right)^{-\frac{2}{3}} m_{q}^{\frac{1}{3}}\,\epsilon\approx 1.0741901\, .
\end{equation}

\subsection{Asymptotic Behavior of the wave function}

The nature of variational calculations is such that they determine energies more accurately than the wave functions.  The Gaussian form of the trial wave function we chose was very convenient and provided for an extremely accurate calculation for the energy. Moreover, with a sufficient number of terms, it provides a rather accurate form for the wave functions over the bulk of the region that contributes substantially to the energy.  However,  it does not capture the correct behavior of the wave function at asymptotically large distances. At large $x$, the drop off of the true solution of Eq.~(\ref{se}) is slower than a Gaussian~\cite{3+1}. Since one principal goal of this paper is to compute the interaction energy of low density baryonic matter -- which is highly sensitive to the long range part of the wave function -- it is important to determine the correct behavior at asymptotic distances.

To proceed, we study Eq.~(\ref{se}) in the limit where $|x|\gg |x'|$. In doing so, we obtain a Schr\H{o}dinger equation that captures the asymptotic behavior of the baryon wave function:
\begin{equation}
\left( -\frac{\partial_{x}^{2}}{2 m_{q}}+\frac{\lambda}{4} |x|\right)\psi_{\rm asy}(x)=\epsilon\, \psi_{\rm asy}(x)\, .
\end{equation}
The general solution is a linear combination of the Airy functions. Taking only the normalizable solution, in units where $\bar{x}=\left (\frac{\lambda}{4} m_{q}\right)^{\frac{1}{3}}x$,  we find $\bar{\psi}_{asy}(\bar{x})=\left ( \frac{\lambda}{4} m_{q}\right)^{-\frac{1}{6}}\psi_{asy}(x)$ and $\bar{\epsilon}\equiv\left (\frac{\lambda}{4} \right)^{-\frac{2}{3}} m_{q}^{\frac{1}{3}}\,\epsilon$, we find that

\begin{equation}
\label{asy}
\bar{\psi}_{\rm asy}(\bar{x})=a \textrm{Ai}(2^{\frac{1}{3}}(|\bar{x}|-\bar{\epsilon}))\, .
\end{equation}
where $a$ is a constant.

The asymptotic form is thus fixed entirely by the determination of the constant $a$.  This can be done numerically provided that there is a region for which $x$ is large enough so that the asymptotic form of the wave function of Eq.~(\ref{asy}) is accurate while simultaneously being small enough that the variational wave function is accurate.  If so, one can match to the two forms in the region of simultaneous validity of the two forms.  It is easy to verify that such a region exists: a plot of the ratio of the two forms of the wave function must have a plateau with a value near unity over an extended region.  Note that the size of such a region of stability depends on the accuracy of the variational wave function.  It was largely for this reason, that we used as many terms as we did in the variational wave function.   Such a stable region was found in the vicinity of $\bar{x}=3$ and the asymptotic wave function takes the form:
\begin{equation}\label{Asyform}\begin{split}
&\bar{\psi}_{\rm asy}(\bar{x})= k  \frac{e^{-\frac{2}{3}[2^{\frac{1}{3}}(|\bar{x}|-\bar{\epsilon})]^{\frac{3}{2}}}}{2\sqrt{\pi} [2^{\frac{1}{3}}(|\bar{x}|-\bar{\epsilon})]^{1/4}}\\
&\textrm{with }k\approx 1.21\textrm{ and }\bar{\epsilon}\approx 1.07\, ,
\end{split}\end{equation}
where the asymptotic form of the Airy function has been used.

\section{Comparison of single baryon with more general approaches }
\label{windingsection}
In this section, we compare our single baryon energy results with numerical results based on the Hamiltonian approach of Refs.~\cite{barak} and~\cite{negele}.  These approaches are more general than ours as they are valid for arbitrary quark masses.  Our approach complements these; while the regime is far more limited, the approach is more transparent and far less reliant on numerical work.   Moreover, there is real value in having the ability to independently check the analysis of Ref.~\cite{barak} in a regime where the results of infinite box size are clearly under control.

Ref.~\cite{barak} analytically finds the most general Hamiltonian for large $N_{c}$ finite density QCD in the baryon sector for arbitrary quark masses in a finite box. The  Hamiltonian for a box of size $L$ is given by~\cite{barak}:
\begin{widetext}
\begin{equation}
\label{hamiltonian}
\begin{split}
\frac{\mathcal{H}}{N_{c}}=\lim_{M\rightarrow\infty}-\frac{i}{2}\int_{0}^{L} dxdy\frac{1}{M}\sum_{j=1}^{M}  \bigg [ \left\{\mathrm{tr} \sigma_{3} \partial_{x} \rho^{j}(x,y)+c.c.+m_{q}\ \mathrm{tr} \sigma_{1} \rho^{j}(x,y)\right\}\delta(x-y)\\
-\frac{\lambda}{8} \frac{1}{M}\sum_{j'=1}^{M}\mathrm{tr}\{\rho^{j}(x,y) \rho^{j'}(y,x)\}\frac{L+(x-y)(e^{-\frac{2\pi i(j-j')}{M}}-1)}{2\sin^{2}(\frac{\pi(j-j')}{M}+i\epsilon)} \bigg ]\, ,
\end{split}
\end{equation}
\end{widetext}
where the the trace is only over Dirac indices.  The Hamiltonian is given in terms of density matrices  $\rho^{j}(x,y)$ but can also be represented in terms of the conjugate space density matrix $P^{Q}(x,y)$ as
\begin{equation}
\label{densitymatrix}
\rho^{j}(x,y)=\sum_{Q\epsilon Z} P^{Q}(x,y)e^{-\frac{2\pi i j Q}{M}}\, .
\end{equation}
The index $Q$ represents integers and $P^{Q}(x,y)$ is defined as:
\begin{equation}
\label{densitymatrix1}
P^{Q}(x,y)=\langle\hspace{1pt}  \langle C[A]|\psi^{\dagger}(y)U^{\dagger}(y,x+QL)\psi(x)|C[A]\rangle\hspace{1pt}\rangle_{A} \, .
\end{equation}
In Eq. (\ref{densitymatrix1}), $\langle \mathcal{O} \rangle_{A}$ represents the average over all possible gauge field configurations of some gauge-invariant quantity $\mathcal{O}$
\begin{equation}
\langle\mathcal{O}\rangle_{A}=\int d[A] \mathcal{O} \ ,
\end{equation}
$U$ is a path-ordered (denoted $\mathcal{P}$) spatial gauge field operator
\begin{equation}
U(x,y)=\mathcal{P}e^{i g \int_{x}^{y} A(x')dx'}\, ,
\end{equation}
and $|C[A]\rangle$ represents appropriately normalized coherent states
\begin{equation}
|C[A]\rangle =e^{\int dx dy C(x,y)g(x,y)}|0\rangle\, .
\end{equation}
Note that $|C[A]\rangle$ has implicit dependence on gauge fields (denoted $A$) via $g(x,y)$, the gauge-invariant generators of the fermionic coherent group \cite{barak}
\begin{equation}
g(x,y)=\psi^{\dagger}(x)U(x,y)\psi(y)\ .
\end{equation} 
$C(x,y)$ are the corresponding weights to $g(x,y)$ and $|0\rangle$ is a color singlet reference state that satisfies the condition $a_{n} |0\rangle$=0, where $a_{n}$ is a member of a set of single particle annihilation operators chosen to ensure that the reference state has the baryon number of interest in the box.

To recap, the diagonal projection onto coherent states defined in Eq. (\ref{densitymatrix1}) is for a particular gauge configuration; however, $P^{Q}$ reflects averaging over all possible gauge configurations. In this averaging, the only non-trivial structure that survives are the windings $Q$ around the periodic box.

Note that $\rho^{j}(y,x)$ has the following properties:
\begin{equation}
\label{densitymatrix2}
\begin{split}
 \rho^{j}(x,y)&=\rho^{j}(y,x)^{*}\\ 
  \int_{0}^{L} \rho^{j}(x,y)\rho^{j}(y,z)dy &=\rho^{j}(x,z)\\ 
  \int_{0}^{L} dx \textrm{ tr}(\rho^{j}(x,x)-\rho^{j}_{vac}(x,x))&=B\, .
\end{split} \end{equation}
Here $\rho^{j}_{vac}(x,x)$ is the density matrix associated with the baryon Dirac sea and gives a  formally divergent contribution to the integral in the absence of regularization. $B$ is the baryon number in the box, which can be written in terms of $b$, the baryon density as $B\equiv bL$.

The baryon Hamiltonian of Eq. (\ref{hamiltonian}) can only be solved numerically. To make the problem numerically tractable, we truncated the theory such that $M$ in Eq.~(\ref{hamiltonian}) is kept finite. This truncation introduces corrections to observables (including the baryon mass) that scale as $\frac{1}{M}$. To obtain the observables in the full theory, the observables are calculated for different values of $M$ and the result in the limit $M\rightarrow\infty$ is obtained by extrapolation.

To formulate the problem numerically, the Hamiltonian is also discretized in space and the continuum results are obtained by extrapolation to the continuum limit. A latticized Hamiltonian using staggered fermions is discussed in detail in Ref. \cite{barak}.  Note that the act of latticizing the theory introduces an ultraviolet cutoff thereby regularizing the theory and rendering the vacuum density matrix,  $\rho^{j}_{vac}(x,x)$, finite.
Here, we only present the form of the density matrix $\rho^{j}(x,y)$ latticized using staggered fermions. These matrices satisfy the properties in Eq.~(\ref{densitymatrix2}) can be written in terms of a finite dimensional basis of orthogonal wave functions ${\phi_{n}^{j}(x)}$ for each $j$ with the indices $n,x=1,2...L_{s}$, where $L_{s}$ is the number of lattice sites. The discretized density matrix is
\begin{equation}
\label{discrete}
\rho^{j}(x,y)=\sum_{n=1}^{B+\frac{L_{s}}{2}}\phi^{j}_{n}(x)\phi^{j}_{n}(y)^{*}\, ,
\end{equation}
with the following orthonormality constraints on the wave functions:
\begin{equation}
\begin{split}
\sum_{x=1}^{L_{s}}\phi^{j}_{n}(x)\phi^{j}_{m}(x)^{*}=\delta_{nm}\, ,\\
\sum_{n=1}^{L_{s}}\phi^{i}_{n}(x)\phi^{j}_{m}(x)^{*}=\delta_{ij}\, .
\end{split}
\end{equation}

The sum from $n=1 \textrm{ to } \frac{L_{s}}{2}$ in Eq.~(\ref{discrete}) is the lattice version of the Dirac trace of vacuum density matrix $\rho_{vac}^{p}(x, x)$ in Eq. (\ref{densitymatrix2}). The integral becomes regularized on the lattice and gives a finite contribution.
 
The Hamiltonian of Eq. (\ref{hamiltonian}) was derived in the axial gauge representation of QCD~\cite{axialQCD}. It is well-known that in the Hamiltonian approach to either QCD or QED, Gauss' law appears as a constraint and not as an equation of motion. Ref.~\cite{axialQCD} explicitly constructed a QCD Hamiltonian, with the Gauss' law constraint built in a manner that only the physical states of the Hilbert space play a role.

Ref.~\cite{barak}, uses the constrained Hamiltonian of Ref.~\cite{axialQCD} and the coherent states approach of Ref.~\cite{yaffe} to solve the Hamiltonian in the baryonic sector, which is the leading order physics in a $\frac{1}{N_c}$ expansion.  This follows from the fact that the coherent states form an overcomplete set so that any physical state can be expressed as an integral over them and that the overlap between coherent states becomes sharply peaked as $N_c \rightarrow \infty$.

The Hamiltonian has a number of interesting features. First, for translationally invariant states, the classical Hamiltonian for the baryons exhibits volume independence and partial volume independence~\cite{barak, unsal}. Second, color charges interact via a linear Coulomb potential of the form $|x-y+QL|$, where $L$ is the size of the box and $Q$ refers to the number of windings around the box. The sign of $Q$ determines the direction of the windings. The presence of these windings is a result of the inability to gauge away all the spatial gauge degrees of freedom in a finite box. Specifically, there are $N_{c}-1$ spatial gluon windings that remain in the axial gauge after the imposition of the Gauss' law constraint.

Ref.~\cite{negele}, on the other hand, formulates the baryon Hamiltonian in infinite volume, where the windings represented by $Q$ are absent; the form of the Hamiltonian is exactly the same as that of Eq.~(\ref{hamiltonian}) except that $Q=0$. However, in constructing the lattice regularized version of the resulting Hartree-Fock equations, Ref.~\cite{negele} assumes that the correct approach is to use a box large enough such that finite volume effects on the baryon structure is small and with lattice spacing small enough that the baryon density can be easily resolved. For this purpose, Ref.~\cite{negele} constructs a lattice potential, which has the form $|x-y|$ in the continuum and infinite volume limits, while continuing to ignore the windings. Thus, the resulting latticized theory of Ref. \cite{negele} is equivalent to the $M=1$ truncation of the Hamiltonian in Eq.~(\ref{hamiltonian}) and, for small boxes, may have significant errors.  One expects these errors to go to zero as the box size goes to infinity.

\subsection{Comparison of Single Baryon Mass and Density Profile}
\begin{figure}[ht]

\centering
\includegraphics[width=3in]{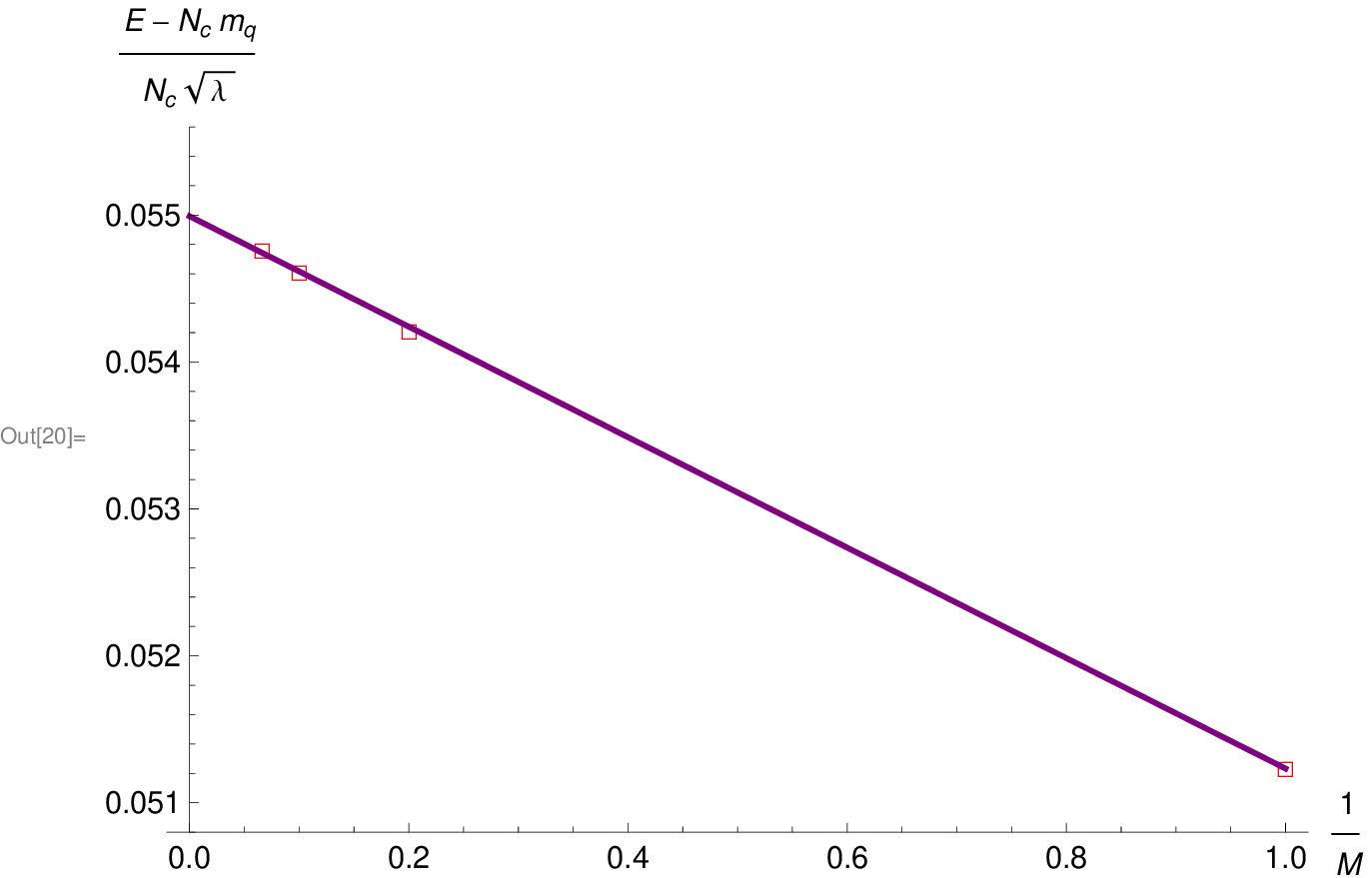}
\caption{Baryon Mass Extrapolation for $\frac{m_{q}}{\sqrt{\lambda}}=100$ in a box of size $L\sqrt{\lambda}=2\sqrt{2\pi}$. Only interaction energies are shown.}
\label{mass100}
\end{figure}

\begin{figure}[ht]
\centering
\includegraphics[width=3in]{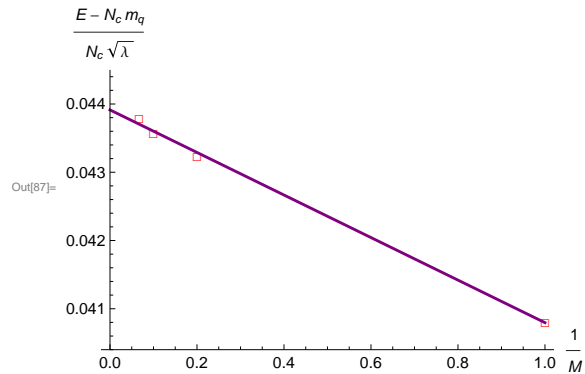}
\caption{Baryon Mass Extrapolation for $\frac{m_{q}}{\sqrt{\lambda}}=200$ in a box of size $L\sqrt{\lambda}=1.6\sqrt{2\pi}$. Only interaction energies are shown.}
\label{mass200}
\end{figure}

We find that in the limit of large quark masses, there is excellent agreement between the variational calculation of section \ref{vari} and a numerical calculation based on the latticized baryon Hamiltonian of Ref.~\cite{barak}. In Figs. \ref{mass100} and \ref{mass200}, we first plot the baryon mass minus the quark mass contribution ({\it i.e.} the interaction energy) as an extrapolation (in $\frac{1}{M}$) for boxes of size $L\sqrt{\lambda}=2\sqrt{2\pi}$ for $\frac{m_{q}}{\sqrt{\lambda}}=100$ and $L\sqrt{\lambda}=1.6\sqrt{2\pi}$ for $\frac{m_{q}}{\sqrt{\lambda}}=200$. Note that the points plotted are the baryon masses in the continuum limit but for different finite values of $M$. We have points for $M=1,5,10,15$.

The baryon masses (including numerical uncertainties) are $\frac{E}{N_{c}\sqrt{\lambda}}= 100.0550\pm 0.0001$ for quark masses of $\frac{m_{q}}{\sqrt{\lambda}}=100$ and $\frac{E}{N_{c}\sqrt{\lambda}}= 200.0440\pm 0.0003$ for quark masses of  $\frac{m_{q}}{\sqrt{\lambda}}=200$. The uncertainties in the baryon masses are associated with taking the continuum and $M\rightarrow \infty$ limits and also the systematic errors due to the numerical nature of the Hartree-Fock calculation. We find that the extrapolated baryon masses are in agreement with the masses from the variational calculation presented in Eq. (\ref{energy}), up to the numerical uncertainties of the calculation; the baryon masses from the variational calculation are $\frac{E_{\rm var}}{N_{c}\sqrt{\lambda}}=100.055098$ and $\frac{E_{\rm var}}{N_{c}\sqrt{\lambda}}=200.043731$ for quark masses of $\frac{m_{q}}{\sqrt{\lambda}}=100$ and $\frac{m_{q}}{\sqrt{\lambda}}=200$ respectively.

\begin{figure}
\centering
\includegraphics[width=3in]{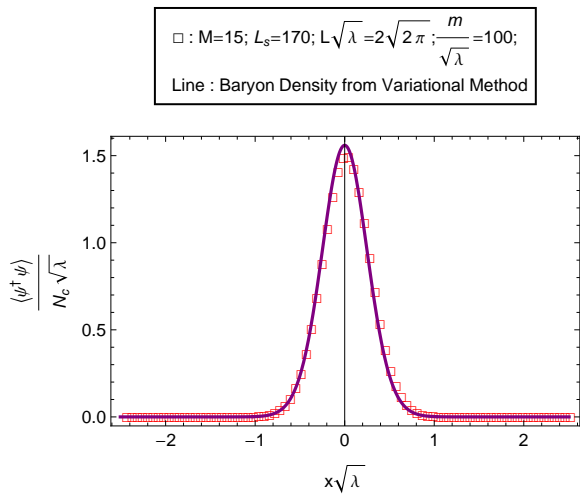}
\caption{Baryon Density for $\frac{m_{q}}{\sqrt{\lambda}}=100$ with $M=15$ and $L_{s}=170$ in a box of size $L\sqrt{\lambda}=2\sqrt{2\pi}$.\label{den1}} 
\end{figure}

\begin{figure}
\centering
\includegraphics[width=3in]{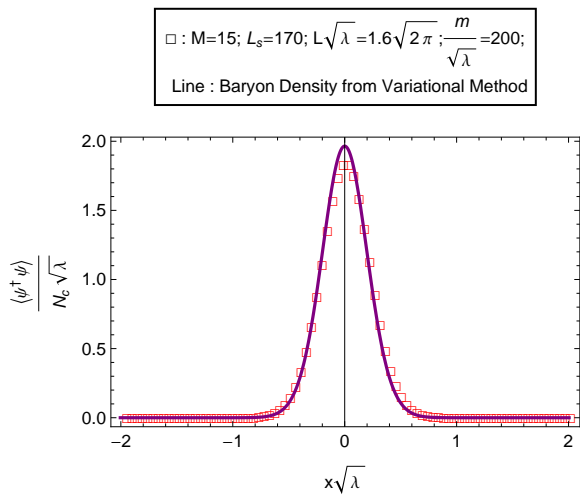}
\caption{Baryon Density for $\frac{m_{q}}{\sqrt{\lambda}}=200$ with $M=15$ and $L_{s}=170$ in a box of size $L\sqrt{\lambda}=1.6\sqrt{2\pi}$.\label{den2}}
\end{figure}

Next, in Figs.~\ref{den1} and \ref{den2}, we plot baryon density profiles for the two calculations with the line representing the result of the variational calculation and the dots representing the result of the numerical calculation of the general finite baryon density Hamiltonian. The densities are plotted for $M=15$ and with the number of lattice points, $L_{s}=170$. The two results agree quite well in spite of the errors associated with discretization of space and truncation in the number of windings.

It is useful to see the extent to which  including the windings  ({\it i.e.} $M \ne 1$) is crucial for getting interaction energies accurately  for system with relatively large-sized boxes.   From the plots in  Figs.~\ref{den1} and \ref{den2} it is clear that the baryons fit well within the boxes in the sense that the baryon density has dropped extraordinarily close to zero well before the edge of the box.  Thus, one might be tempted to conclude that these boxes are big enough so that the effects of winding should be quite small.  However, this is not the case. Comparing the interaction energy of the $M=1$ baryon mass (which we label by $E_{h}$) with the interaction energy after $M\rightarrow \infty$ extrapolation (labelled $E_{H}$) using the quantity $\frac{E_{H}-E_{h}}{\frac{1}{2}(E_{H}+E_{h})}$, which measures the relative difference in the interaction pieces of the extrapolated and the $M=1$ baryon masses. We also quote the error in the $M=1$ baryon mass, which comes from systematic errors associated with the numerical nature of the Hartree-Fock calculation and the continuum extrapolation. For a quark mass of $\frac{m_{q}}{\sqrt{\lambda}}=100$, the $M=1$ baryon mass is $\frac{E}{N_{c}\sqrt{\lambda}}=100.05124\pm 0.00002$ and for quark masses of $\frac{m_{q}}{\sqrt{\lambda}}=200$, the $M=1$ baryon mass is $\frac{E}{N_{c}\sqrt{\lambda}}=200.04081\pm 0.00002$.
Thus,  we find that the relative difference in interaction energy between the $M=1$ result is approximately 7 per cent, which is quite significant given the degree to which the box fully contains the baryon.

The large difference between the $M=1$ and $M=\infty$ results show the practical importance of the windings in realistic calculations even for very large boxes.  Of course,  the effect of the winding is expected to be large for {\it small} boxes and was introduced largely to explain the behavior of the system in small boxes including large $N_c$ volume independence \cite{barak}.   To some extent, the large effect of the windings on the boxes computed here is rather surprising.  Indeed, they are large enough that one might question whether there is something wrong with the formalism in \cite{barak}.  However, the fact that the results extrapolated to $M$ going to infinity agree so well with the results derived with the formalism of this paper, which are known to be valid for heavy quark masses, shows that this is not the case.

\section{Baryonic matter}

\subsection{Introduction to the Multi-baryon Problem}

Now that we have established that Witten's heavy-quark formulation for single baryons is consistent with results based on the Hamiltonian approach of Ref. \cite{barak}, we turn our attention to baryonic matter. It is believed that baryonic matter forms a crystal at large $N_c$: the baryons are heavy and have a tendency lie in equilibrium at the potential wells created by the other baryons. 

We consider infinite nuclear matter at low enough densities such that only interbaryon interactions between nearest neighbors are relevant. 

Here, we assume that the many-baryon sector consists of distinct color-singlet clusters of $N_c$ quarks with identical wave functions, as in Witten's discussion of baryon scattering~\cite{witten}. An analogous assumption was made in Ref.~\cite{3+1} for heavy quark, large $N_{c}$ QCD in 3+1 dimensions. This assumption is highly plausible but as noted in Ref.~\cite{3+1}, it has not been rigorously established. A possible concern is that hidden color states could contribute at leading order. Thus, it is useful to test this assumption in a 1+1 dimensional setting.  We can do that by comparing the results based on this assumption using our heavy quark formalism with numerical results using the methods of Ref.~\cite{barak}, which are known to be valid at large $N_c$ generally and do not involve any additional assumptions.
 
Let us begin by deriving the interaction energy per baryon in a 1+1 dimensional crystal at low density  using the heavy quark non-relativistic formalism.  The energy per unit-cell of the low-density baryon crystal in terms of a reference baryon is: 
\begin{widetext}
\begin{equation} 
\label{bothterms}
\begin{split}
\frac{\bar{E}}{N_c}&=\frac{1}{2}\int d\bar{x}\,(\partial_{\bar{x}} \bar{\Psi}_i(\bar{x}))^2+\frac{1}{2}\int d\bar{x}\,d\bar{y}\,|\bar{\Psi}_i(\bar{x})|^2|\bar{\Psi}_i(\bar{y})|^2|\bar{x}-\bar{y}|\\
    &+\frac{1}{2}\int\,d\bar{x}\,d\bar{y}\,\bar{\Psi}_i(\bar{x})\bar{\Psi}_{i-1}(\bar{y})\bar{\Psi}_{i-1}(\bar{x})\bar{\Psi}_{i}(\bar{y})|\bar{x}-\bar{y}|+\frac{1}{2}\int\,d\bar{x}\,d\bar{y}\,\bar{\Psi}_i(\bar{x})\bar{\Psi}_{i+1}(\bar{y})\bar{\Psi}_{i+1}(\bar{x})\bar{\Psi}_{i}(\bar{y})|\bar{x}-\bar{y}| \, ,
\end{split}
\end{equation}
\end{widetext}
where $\bar{\Psi}_i(\bar{x})$ is the wave function of the reference baryon, $\bar{\Psi}_{i\pm 1}(\bar{x})$ are the wave functions of the nearest-neighbor baryons, the first term is the contribution to the intrabaryon kinetic energy, the second term is the contribution to the intrabaryon potential energy, and the last two terms are the contributions to the interbaryon energy. In this expression both $\lambda$ and $m_{q}$ have been scaled out of the energy density using the following change of variables:
\begin{equation}
\begin{split}
\label{units}
&\bar{x}=\left( \frac{\lambda}{4} m_{q}\right)^{\frac{1}{3}}x\\
&\bar{\psi}=\left( \frac{\lambda}{4} m_{q}\right)^{-\frac{1}{6}}\psi\\
&\bar{E}=\left( \frac{\lambda}{4}\right)^{-\frac{2}{3}} m_{q}^{\frac{1}{3}}E\, .
\end{split}
\end{equation}
By assumption, the density is sufficiently small so that the effect of interactions with next-to-nearest neighbors is negligible.

The wave function for the baryon crystal is constrained by the Pauli exclusion principle, which precludes two fermions from simultaneously occupying the same position. This implies  that the wave function of a reference baryon state $\bar{\Psi}(\bar{x})$  must be completely orthogonal to the the wave function of the neighboring  baryon states $\bar{\Psi} (\bar{x}+\bar{d})$ and $\bar{\Psi} (\bar{x}-\bar{d})$. 
\begin{equation}\label{wavefuntionform}\begin{split}
&\bar{\Psi}(\bar{x})=(1+\kappa)\bar{\psi}_{0}(\bar{x})-\alpha \bar{\phi}(\bar{x}-\bar{d})-\alpha\bar{\phi}(\bar{x}+\bar{d})\\
&\text{with}\ \int d\bar{x}\,|\bar{\Psi} (\bar{x})|^{2}=1\\
&\text{and}\ \int d\bar{x}\,|\bar{\Psi} (\bar{x})||\bar{\Psi} (\bar{x} \pm \bar{d})|=0\, ,
\end{split}\end{equation}
where $\bar{\phi}(\bar{x})$ is some normalized wave function.
Here, we have assumed that densities are low enough that the reference baryon wave function is dominated by the single baryon wave function $\bar{\psi}(\bar{x})$ except for corrections centered around $\bar{x}=\pm \bar{d}$. The choice of $\kappa$ and $\alpha$ is determined by the orthogonality constraint of the neighboring baryons and by normalization of the baryon wave function. We find $\kappa$ and $\alpha$ in terms of $\gamma\equiv\int d\bar{x}\, \bar{\psi}_{0}(\bar{x})\bar{\phi} (\bar{x})$, which characterizes the overlap at the same spatial points of the neighboring baryon wave functions and the quantity $\mathcal{A}'\equiv\int d\bar{x}\,\bar{\psi}_{0}(\bar{x}) \bar{\phi} (\bar{x}+\bar{d})$, which characterizes the orthogonality of the neighboring wave functions:
\begin{equation}\begin{split}
\label{overlap}
&\alpha=\frac{\mathcal{A}}{2\gamma}\ \text{and}\ \kappa=\frac{\mathcal{A} \mathcal{A}'}{\gamma}-\frac{\mathcal{A}^{2}}{4\gamma^{2}}\\
&\textrm{where } \mathcal{A} \equiv \int d\bar{x}\,\bar{\psi}_{0}(\bar{x})\bar{\psi}_{0}(\bar{x} \pm \bar{d}) \, .
\end{split}\end{equation}
We have disregarded $\mathcal{O}(\mathcal{A}^{3})$ effects because $\mathcal{A}$ becomes exponentially small with decreasing density and in the low density regime, the higher order effects are parametrically small. Effectively, we are writing the interaction energy in the form $\mathcal{A}^2 f\left (\frac{w}{d} \right )$ (where $w$ is the characteristic width of a baryon and $d$ is the separation of neighboring baryons) by developing an appropriate power expansion for $f\left (\frac{w}{d} \right )$.  

To proceed, we first  assume that $\bar{\phi} (\bar{x})=\bar{\psi}_{0}(\bar{x})$, which appears to be the most efficient way to orthogonalize neighboring crystal wave functions.  We will show {\it a posteriori } that this assumption is correct in the sense that it will minimize the total energy up to corrections which are parametrically small at low density. The orthogonality constraint in doing so gives
\begin{equation}
\int d\bar{x}\,\bar{\Psi} (\bar{x}) \bar{\Psi}(\bar{x}+\bar{d})=-\mathcal{A}(\bar{d})+\mathcal{O}(\mathcal{A}^{3})\ .
\end{equation}

As can be seen from the explicit form of $\mathcal{A}(\bar{d})$, which we calculate in the following subsection, for parametrically low densities $\mathcal{A}(\bar{d})\sim e^{-\bar{d}^{\frac{3}{2}}}$. Hence, it is plausible that in the low density regime, where only nearest-neighbor interactions are relevant, the choice of $\bar{\phi} (\bar{x})=\bar{\psi}_{0}(\bar{x})$ is indeed the correct one. In the section \ref{Selfcons}, we introduce corrections such that $\bar{\phi}(\bar{x})=\eta(\bar{\psi}_{0}(\bar{x})+\delta \bar{\Delta}(\bar{x}))$, where $\bar{\Delta}$ is a normalized wavefunction and argue that contributions to energy at $\mathcal{O}(\delta)$ do not contribute at the order to which we work.   

\subsection{Calculation of overlap function $\mathcal{A}$}

Before proceeding with the calculation of energy, it is useful to calculate the overlap function $\mathcal{A}$, which is defined in Eq. (\ref{overlap}). Most of the contribution to $\mathcal{A}$ comes from the localized baryon density at $\bar{x} =\frac{\bar{d}}{2}$. In the low density regime, the contributions to $\mathcal{A}$, localized at $\bar{x} =0$ and $\bar{x} =\bar{d}$ are exponentially suppressed. The contribution to the center comes from the tails of neighboring baryons, which have the form of Airy functions.
Using the asymptotic form of Airy functions, we get

\begin{equation} \label{A}
\mathcal{A}=\int d\bar{x}\, \frac{k^2 e^{-\frac{1}{3} (2 \bar{d} -2 \bar{\epsilon} -2 \bar{x} )^{\frac{3}{2}}-\frac{1}{3} (-2 \bar{\epsilon} +2 \bar{x} )^{\frac{3}{2}}}}{2^{\frac{5}{3}} \pi  (2 \bar{d} -2 \bar{\epsilon} -2 \bar{x} )^{\frac{1}{4}} (-2 \bar{\epsilon} +2 \bar{x} )^{\frac{1}{4}}}\, .
\end{equation}
where $k\approx 1.21$ is the strength of the asymptotic wave function of Eq.~(\ref{Asyform}).

The function in the exponent has a  sharp minima at $\bar{x}=\frac{\bar{d}}{2}$ and hence  the integrand contributing to $\mathcal{A}$ has a maxima at $\bar{x}=\frac{\bar{d}}{2}$. We use the standard steepest descent method to calculate the integral, which involves series expanding the exponent around the turning point, evaluating the denominator of $\mathcal{A}$ at $\bar{x}=\frac{\bar{d}}{2}$ and performing the integral. This gives
\begin{equation}
\mathcal{A}(\bar{d})=\frac{k^2 e^{-\frac{2}{3} (\bar{d} -2 \bar{\epsilon} )^{\frac{3}{2}}}}{2^{\frac{5}{3}}\sqrt{\pi} (\bar{d} -2 \bar{\epsilon} )^{\frac{1}{4}}}\, ,
\end{equation}
with a leading order correction term that is proportional to $\frac{e^{-\frac{2}{3} (\bar{d} -2 \bar{\epsilon} )^{3/2}}}{(\bar{d} -2 \bar{\epsilon} )^{7/4} }$. The correction is suppressed for parametrically large distances, which is the relevant regime for this paper. 

\subsection{Energy at low density}
Here, we calculate the interaction energy per unit cell of a low density baryon crystal up to $\mathcal{O}(\mathcal{A}^2)$ and leading order in $\mathcal{O}\left(\frac{w}{d}\right)$ using units of Eq. (\ref{units}). Note that $w$ is the characteristic width of an isolated baryon and $d$ is the interbaryon separation.  We divide the energy into two parts: the first line of Eq.~(\ref{bothterms}), which label the intrabaryon contributions, and the second line, which label the interbaryon contribution.  The intrabaryon contributions contain a contribution from the kinetic energy and another from a potential energy. Note that the intrabaryon energy is not the same as in the free space baryon since the wave function has been modified due to the interactions with the other baryons.

\subsubsection{Interbaryon Energy}
The contribution to the interbaryon energy under the assumption that only nearest neighbor interactions are relevant is presented in Eq.~(\ref{interb}).  
\begin{widetext}
\begin{equation}
\label{interb}
\frac{\bar{E}^{\rm interbaryon}}{N_{c}}=\frac{1}{2}\int\,d\bar{x}\,d\bar{y}\,\bar{\Psi}(\bar{x})\bar{\Psi}(\bar{x}+\bar{d})\bar{\Psi}(\bar{y})\bar{\Psi}(\bar{y}+\bar{d})|\bar{x}-\bar{y}|+\frac{1}{2}\int\,d\bar{x}\,d\bar{y}\,\bar{\Psi}(\bar{x})\bar{\Psi}(\bar{x}-\bar{d})\bar{\Psi}(\bar{y})\bar{\Psi}(\bar{y}-\bar{d})|\bar{x}-\bar{y}|\ ,
\end{equation}

\begin{equation}\begin{split}
\label{interdensity}
&\textrm{where } \bar{\Psi}(\bar{x})\bar{\Psi}(\bar{x}\pm\bar{d})=(1+\kappa)^2\bar{\psi}_{0}(\bar{x})\bar{\psi} _{0}(\bar{x}\pm \bar{d})-\alpha(1+\kappa)\bar{\psi} _{0}(\bar{x})\bar{\psi}_{0} (\bar{x})-\alpha(1+\kappa)\bar{\psi} _{0}(\bar{x}\pm \bar{d})\bar{\psi}_{0} (\bar{x}\pm \bar{d})\\
&\hspace{2.95cm}+\alpha^{2}(\bar{\psi}_{0} (\bar{x}\pm \bar{d})\bar{\psi}_{0} (\bar{x}\pm 2\bar{d})+\bar{\psi}_{0} (\bar{x})\bar{\psi}_{0} (\bar{x}\pm \bar{d})+\bar{\psi}_{0} (\bar{x})\bar{\psi}_{0} (\bar{x}\mp \bar{d}))\ .
\end{split}\end{equation}
\end{widetext}
The two terms in the equation correspond to the interbaryon energies in a unit cell associated with two nearest neighbors of a reference baryon state $\bar{\Psi}(\bar{x})$ in the crystal. Note that all terms in the interbaryon energy in Eq. (\ref{interb}) are of the form $\int d\bar{x}\,d\bar{y}\,\rho(\bar{x})\rho(\bar{y})|\bar{x}-\bar{y}|$, which is of the same structure as the Coloumb energy associated with a one-dimensional charge density $\rho(\bar{x})$. Thus, for example, the first term in Eq.~(\ref{interb}) has an effective charge density $\bar{\Psi}(\bar{x})\bar{\Psi}(\bar{x}+\bar{d})$.   Moreover, note that in one spatial dimension, unlike in three spatial dimensions, the Coloumb energy of a charge distribution grows with the size of a system.

In Eq.~(\ref{interdensity}), we present the effective charge density that determines the interbaryon energy in terms of the single baryon wavefunction $\bar{\psi}_{0}(x)$, $\alpha$ and $\kappa$. Equation (\ref{overlap}) completely determines the two parameters. Since $\bar{\phi}=\bar{\psi}_{0}$ for a low-density baryon crystal,  we have $\gamma=1$ and $\mathcal{A}'=\mathcal{A}$. Therefore, $\alpha=\frac{\mathcal{A}}{2}$ and $\kappa=\frac{3}{4}\mathcal{A}^{2}$.

\begin{figure}
\centering
\includegraphics[width=3in]{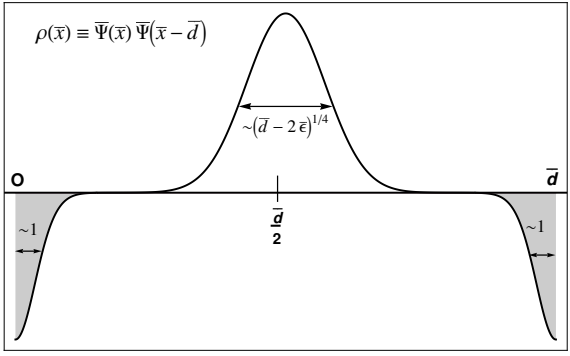}
\caption{Schematic plot of the effective charge density $\Psi(\bar{x})\Psi(\bar{x}-\bar{d})$ associated with the interbaryon energy including the characteristic width of the relevant baryon densities in units of Eq.~(\ref{units}).  The magnitude of the effective charge in each region is $\mathcal{A}$.  The total effective charges are $\mathcal{A}$ in the peak at $\bar{x}=\frac{\bar{d}}{2}$ and $-\frac{\mathcal{A}}{2}$ each in the peaks at $\bar{x}=0\ , \bar{d}$}
\label{schematicinter}
\end{figure}

To proceed, we note that the effective charge density relevant for the interbaryon contribution is localized.  In Fig.~\ref{schematicinter}, we show a schematic  effective charge density plot characterizing the nature of $\bar{\Psi}(\bar{x})\bar{\Psi}(\bar{x}-\bar{d})$ in Eq. (\ref{interdensity}) upto $\mathcal{O}(\mathcal{A}^{2})$. Note that the densities are sharply peaked at $\bar{x}=0\ ,\frac{\bar{d}}{2}\ ,\bar{d}$. These peaks in Fig.~\ref{schematicinter} correspond to the first three terms of Eq.~(\ref{interdensity}). The smaller peaks at $\bar{x}=0$ and $\bar{x}=\bar{d}$ correspond to the second and the third terms of Eq. (\ref{interdensity}) respectively. Since the isolated baryon wave funtion, $\bar{\psi}_{0}(\bar{x})$, is normalized, the terms each have a total baryon charge of $-\frac{\mathcal{A}}{2}$. Furthermore, the width (in units of Eq.~(\ref{units})) of the peak at $\bar{x}=0$, which is proportional to the charge density, $|\bar{\psi}_{0}(\bar{x})|^{2}$, is of $\mathcal{O}(1)$ and does not scale with $\bar{d}$ as can be seen from the results of Eqs.~(\ref{three}), (\ref{six}) and (\ref{eleven}). The same properties are true for the charge density centered at $\bar{x}=\bar{d}$. However, the effective charge density centered around $\bar{x}=\frac{\bar{d}}{2}$ behaves differently. For large interbaryon separation $\bar{d}$, the width (in units of Eq.~(\ref{units})) scales as $(\bar{d}-2\bar{\epsilon})^{\frac{1}{4}}$, which can be easily calculated from the standard steepest descent calculation in the integrand of Eq.~(\ref{overlap}). Also note that the total effective charge in the center peak is $\mathcal{A}$ for parametrically small baryon densities.  Although the width of the central peak depends on $\bar{d}$ (i.e. the baryon density), it is much smaller than the interparticle spacing at low density.  Thus in the low-density regime, these three regions of charge are well separated from each other with a spacing much larger then the width of each lump.

To calculate the leading contribution interbaryon energy per unit cell of a baryon crystal in the low-density regime, the effective charge densities can be approximated as delta functions with the appropriate baryon charges.  The leading correction is  the fact that the spatial extent of the charge density centered around $\bar{x}=\frac{\bar{d}}{2}$ grows with $\bar{d}$. The size of this correction can be calculated using the approximation to the asymptotic form of the Airy function that was used to calculate the overlap integral $\mathcal{A}$ in Eq.~(\ref{interb}). This integral can be calculated easily using the method of steepest descent. 

For parametrically low baryon densities, the total interbaryon energy becomes:
\begin{equation}
\frac{\bar{E}^{\textrm{interbaryon}}}{N_{c}}=-\frac{\mathcal{A}^{2}\bar{d}}{2}+\frac{\Delta \bar{E}^{\rm interbaryon}}{N_{c}}\ ,
\end{equation}
where $\frac{\Delta \bar{E}^{\rm interbaryon}}{N_{c}}$ is the correction to the interbaryon energy associated with the fact that the center peak scales with $\bar{d}$:
\begin{equation}
\frac{\Delta \bar{E}^{\rm interbaryon}}{N_{c}}=\sqrt{\frac{2}{\pi}}(\bar{d}-2\bar{\epsilon} )^{\frac{1}{4}}\mathcal{A}^{2}(\bar{d})\, .
\end{equation}
The interbaryon energy above has the form -- charge squared times a width -- which is as expected. Also, note that the energy contributions due to the finite extent of the two regions of effective charge at the the ends is $\mathcal{O}(\bar{d}^0)$ and hence is parametrically smaller at large $\bar{d}$.

\begin{figure}
\centering
\includegraphics[width=3in]{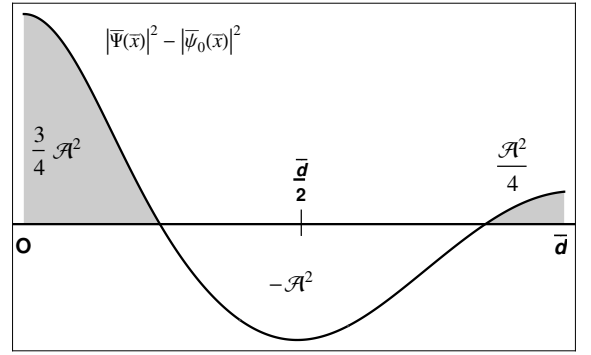}
\caption{Schematic plot of $|\Psi(\bar{x})|^{2}-|\bar{\psi}_{0}(\bar{x})|^{2}$, the effective charge density minus the single baryon charge density for $0<\bar{x}<\bar{d}$ including the total baryon charge in each region.}
\label{schematicintra}
\end{figure}

\subsubsection{Intrabaryon Energy: Potential Contribution}

In this subsection and the next, we calculate intrabaryon energy for the low-density baryon crystal at $\mathcal{O}\left(\frac{w}{d}\right)$. There are two contributions to this, the potential and the kinetic.

The potential contributions to the intrabaryon energy of the baryon crystal comes from Eq.~(\ref{intra}). The effective charge density $|\bar{\Psi}(\bar{x})|^{2}$ is presented in Eq.~(\ref{intradensity}). Additionally, we present in Fig.~\ref{schematicinter} a schematic plot of $|\bar{\Psi}(\bar{x})|^{2}-|\bar{\psi}_{0}(\bar{x})|^{2}$, the shift in the effective baryon charge density from the free case,  $|\bar{\Psi}(\bar{x})|^{2}$, minus the isolated baryon density, $|\bar{\psi}_{0}(\bar{x})|^{2}$, to emphasize the various charge enhancements. This quantity is even around $\bar{x}=0$ and hence we only plot the region $0<\bar{x}<\bar{d}$.

\begin{widetext}
\begin{equation}
\label{intra}
\frac{\bar{E}^{\textrm{intrabaryon}}_{\textrm{potential}}}{N_{c}}=\frac{1}{2}\int d\bar{x}\,d\bar{y}\,|\bar{\Psi}(\bar{x} )|^{2}|\bar{\Psi}(\bar{y} )|^2|\bar{x} -\bar{y}|\ ,
\end{equation}
\begin{equation}\begin{split}
\label{intradensity}
&\textrm{where\ }|\Psi(\bar{x} )|^{2}=(1+\kappa)^{2}\bar{\psi}_{0}(\bar{x} )^2+\alpha^{2}(\bar{\psi}_{0} (\bar{x} +\bar{d} )^{2}+\bar{\psi}_{0} (\bar{x} -\bar{d} )^{2})-2(1+\kappa)\alpha(\bar{\psi} _{0}(\bar{x} )\bar{\psi}_{0} (\bar{x} +\bar{d} )+\bar{\psi}_{0}(\bar{x} )\bar{\psi}_{0} (\bar{x} -\bar{d} ))\\
&\hspace{2cm}+\alpha^{2}(\bar{\psi}_{0} (\bar{x} +\bar{d} )\bar{\psi}_{0} (\bar{x} -\bar{d}))\, .
\end{split}\end{equation}
\end{widetext}

To proceed, we calculate the effective charge density, $|\bar{\Psi}(\bar{x})|^{2}$, upto $\mathcal{O}(\mathcal{A}^{2})$ to obtain the intrabaryon energy at the same order. Note that in Eq. (\ref{intradensity}), the first term, $|\bar{\psi}_{0}(\bar{x})|^{2}$, of the effective charge density is centered around $\bar{x}=0$, the second term proportional to $|\bar{\psi}_{0}(\bar{x}\mp \bar{d})|^{2}$ is centered around $\bar{x}=\pm \bar{d}$, and the third term, which is proportional to $\bar{\psi}_{0}(\bar{x})\bar{\psi}_{0}(\bar{x}\mp \bar{d})$, is centered around $\bar{x}=\pm \bar{d}$.

Furthermore, at $\mathcal{O}(\mathcal{A}^{2})$ the intrabaryon energy contributions for the baryon crystal come from a cross-term between the effective charge from the unperturbed baryon, $\bar{\psi}_{0}(\bar{x})$, and the effective charge due to the perturbation induced by the near-neighbor baryons.  To the extent that $\bar{d}$ is large, the original charge is well-localized at the origin, it creates an effective electric field which is constant for $|\frac{x}{w}| \gg 1$ (or a potential linear in $\bar{d}$).  With a linear field dependence, the potential energy is sensitive to the center of charge the effective charge due to the perturbation.  Thus, we can  approximate the charge densities in Eq.~(\ref{intradensity}) as localized delta functions with the appropriate strengths, which are associated with the total baryon charges of each of the terms.  Now for the case of the first term, which has a strength of $1+\frac{3}{2}\mathcal{A}^{2}$ and is localized at the origin, the delta function approximation is problematic, since it is not in the region where the field from the unperturbed effective charge is linear.  However, the potential energy from this term is of $\mathcal{O}(\mathcal{A}^{2}\bar{w})$, while the leading behavior scales as  $\mathcal{A}^{2}\bar{d}$.  Thus, this term can be neglected as being parametrically small.  The contributions from the other two terms yield,

\begin{equation}
\frac{\bar{E}_{\textrm{potential}}^{\textrm{intrabaryon}}}{N_{c}}=-\frac{\mathcal{A}^2 \bar{d} }{2}\, .
\end{equation}

As noted above, there are corrections to this of relative order $\frac{w}{d}$ from the perturbation in the effective charge due to $\kappa$.  There are other corrections due to the fact that unperturbed effective charge has a spread and hence the effective field is not precisely constant.  These corrections are exponentially small in $\bar{d}$.

\subsubsection{Intrabaryon Energy: Kinetic Contribution}
The kinetic energy contribution to the intrabaryon energy at leading order in $\frac{w}{d}$ is presented in Eq. (\ref{kin}). Note that the first three terms are independent of the interbaryon spacing $\bar{d}$, whereas the remaining terms are not; in the low-density regime, the contribution from these terms independent of $\bar{d}$ are parametrically small. Furthermore, we ignore the $\mathcal{O}(1)$ term because it is the contribution to the single baryon kinetic energy and not the crystal interaction energy. We also ignore the last term because it is not nearest neighbor.
\begin{widetext}
\begin{equation}\begin{split}
\label{kin}
\frac{\bar{E}_{\textrm{kinetic}}^{\textrm{intrabaryon}}}{N_{c}}=\frac{1}{2}\int d\bar{x}\,(\partial_{\bar{x} }\bar{\Psi} (\bar{x} ))^{2}
&=\frac{1}{2}(1+\kappa)^2\int d\bar{x}\,(\partial_{\bar{x} } \bar{\psi}_{0}(\bar{x}))^{2}+\frac{1}{2}\alpha^2\int d\bar{x}\,(\partial_{\bar{x} }\bar{\psi}_{0} (\bar{x} +\bar{d} ))^{2}+\frac{1}{2}\alpha^2\int d\bar{x}\,(\partial_{\bar{x} } \bar{\psi}_{0} (\bar{x} -\bar{d} ))^{2}\\
&-\alpha(1+\kappa)\int d\bar{x}\,(\partial_{\bar{x} }\bar{\psi}_{0}(\bar{x} ))(\partial_{\bar{x}}\bar{\psi}_{0} (\bar{x} +\bar{d} ))-\alpha(1+\kappa)\int d\bar{x}\, (\partial_{\bar{x} } \bar{\psi}_{0}(\bar{x} ))(\partial_{\bar{x} }\bar{\psi}_{0} (\bar{x} -\bar{d} ))\\
&+\alpha^{2}\int d\bar{x}\,(\partial_{\bar{x} } \bar{\psi}_{0} (\bar{x} +\bar{d} ))(\partial_{\bar{x} } \bar{\psi}_{0} (\bar{x} -\bar{d} ))
\end{split}\end{equation}
\end{widetext}
The remaining terms are trivially calculated using integration by parts, the asymptotic form of the Airy function in Eq.~(\ref{asy}) and the steepest descent method to evaluate the integrals $\int d\bar{x}\,(\partial_{\bar{x} }\bar{\psi}_{0}(\bar{x} ))(\partial_{\bar{x}}\bar{\psi}_{0} (\bar{x} \mp \bar{d} ))$, where the integrand is sharply peak around $\bar{x} =\pm\frac{\bar{d}}{2}$.

The kinetic energy contribution to the interaction energy then becomes 
\begin{equation}
\frac{\bar{E}_{\textrm{kinetic}}^{\textrm{intrabaryon}}}{N_{c}}=\mathcal{A}^2\bar{d} \ .
\end{equation}
\subsubsection{The interaction energy}

Finally, putting all the interaction energies together, the interaction  energy per unit cell of a low-density baryon crystal in the heavy quark mass limit up to $\mathcal{O}(\mathcal{A}^{2})$ and to  leading order in $\bar{d}$ is:

\begin{equation}\begin{split}
\label{intenergy}
\frac{\bar{E}^{\textrm{interaction}}}
{N_{c}}&=\frac{\bar{E}^{\rm interbaryon}}{N_{c}}+\frac{\bar{E}_{\rm potential}^{\rm intrabaryon}}{N_{c}}+\frac{\bar{E}_{\rm kinetic}^{\rm intrabaryon}}{N_{c}}\\&=\sqrt{\frac{2}{\pi}}(\bar{d} -2\bar{\epsilon} )^{\frac{1}{4}}\mathcal{A}^{2}(\bar{d})\, .
\end{split}\end{equation}

Equation (\ref{intenergy}) contains as a factor $(\bar{d} -2\bar{\epsilon} )^{1/4}$.  We are  interested in the limit of large $\bar{d}$. So one could, in principle, expand $(\bar{d} -2\bar{\epsilon} )^{\frac{1}{4}}$ as $\bar{d}^{\frac{1}{4}}$ times a series in powers of $\frac{\bar{\epsilon}}{\bar{d}}$. Moreover, since we are working in a regime, where $\bar{d}$ is parametrically large, all of the subleading terms in the series can be dropped. Hence, it may seem appropriate to simply use  $\bar{d}^{\frac{1}{4}}$. However, it is actually prudent to keep all the terms in powers of $\frac{\bar{\epsilon}}{\bar{d}}$.  The reason is that there is a factor of $(\bar{d} -2\bar{\epsilon} )^{-\frac{1}{2}}$ coming from Eq.~(\ref{A}).  This means that our approximate expression for the interaction energy in Eq.~(\ref{intenergy}) breaks down as $\bar{d} \rightarrow 2\bar{\epsilon}$, yielding a diverging energy.  This divergence goes as  $(\bar{d} -2\bar{\epsilon} )^{-\frac{1}{2}}$ but if we replace $(\bar{d} -2\bar{\epsilon} )^{\frac{1}{4}}$ by $\bar{d}^{\frac{1}{4}}$ in the numerator, it diverges much more slowly -- as $(\bar{d} -2\bar{\epsilon} )^{-\frac{1}{4}}$.  This slower rate of divergence suggests that the expression will remain accurate for larger crystal densities.

\subsection{Self consistency}
\label{Selfcons}
Up until this point, we have {\it assumed} that the wave function for a baryon in unit cell was given by Eq.~(\ref{wavefuntionform}) with $\bar{\phi}=\bar{\psi}_0$.  Such a choice seemed natural in that it orthogonalizes the wave functions associated with different unit cells in a very efficient manner.  It  turns out that this ans\"atz can be  shown to be correct: the corrections to it yield changes in the interaction energy that are parametrically smaller than the result given in Eq.~(\ref{intenergy}).

To show this, we introduce corrections to $\bar{\phi}(\bar{x})$ of the form 
\begin{equation}
\label{correction}
\bar{\phi}(\bar{x})=\eta(\bar{\psi}_{0}(\bar{x})+\delta \bar{\Delta}(\bar{x}))\, ,
\end{equation}
where $\delta$ is taken to be a small parameter, $\bar{\Delta}(\bar{x})$ is an arbitrary wave function and $\eta$ is a normalization constant ensuring that $\bar{\phi}(\bar{x})$ is normalized.  Without loss of generality, one can always take take $\bar{\Delta}(\bar{x})$ to be normalized and orthogonal to both $\bar{\psi} (\bar{x})$ and $\bar{\psi} (\bar{x} \pm \bar{d})$.   Any component in $\bar{\Delta}(\bar{x})$,  which are not orthogonal to $\bar{\psi} (\bar{x})$ and $\bar{\psi} (\bar{x} \pm \bar{d})$ simply serves to renormalize $\eta$ and $\kappa$ without altering the wave function.

Our ans\"atz that  $\bar{\phi}=\bar{\psi}_0$ is self-consistently correct if it minimizes the energy up to corrections, which are parametrically small.  This is true provided that for any choice of $\bar{\Delta}$,
\begin{equation} \label{selfconcond}
\frac{1}{N_c} \left . \frac{d \, \bar{E}^{\rm interaction}} {d \, \delta}\right |_{\delta=0} \sim 
{\cal{O}} (\mathcal{A}^{2} \bar{d}^{0})\, ,
\end{equation}
 since the leading term in the energy is parametrically $ {\cal{O}} (\mathcal{A}^2 \bar{d}^{\frac{1}{4}})$.

To proceed, one simply inserts the form in Eq.~(\ref{correction}) into Eqs.~(\ref{wavefuntionform}) and (\ref{bothterms}) and expands to first order in $\delta$.  Our self consistency condition will be satisfied if the terms of $\mathcal{O}(\delta)$ are parametrically of $\mathcal{O}(\mathcal{A}^2\bar{d}^{0})$.  There are a total of 31 terms in the interaction energy at order $\delta$ -- eight coming from the intrabaryon potential contribution, 21 from the interbaryon energy and two from the interbaryon kinetic contribution.  Of these, four terms -- two from the intrabaryon potential contribution and the two from the intrabaryon kinetic contribution -- are each of $ {\cal{O}} (\mathcal{A}^1)$, the other 27 terms are all of ${\cal{O}} (\mathcal{A}^2)$.  It is straightforward to show that while each of the four terms of $ {\cal{O}} (\mathcal{A}^1)$ is nonzero, their sum is exactly zero.  

Finally, there are $\mathcal{O}(\mathcal{A}^{2})$ terms from the interbaryon and intrabaryon energies.  It can be shown on a term-by-term basis that these terms all contribute at $ {\cal{O}} (\mathcal{A}^2\bar{d}^{0})$ or less.  In doing so, the orthogonality of $\bar{\Delta}(\bar{x}))$ with both $\bar{\psi}_{0}(\bar{x})$ and $\bar{\psi}_0(\bar{x}\pm{\bar{d}})$ play a critical role since this implies that the integrated effective charge density due to cross-terms between $\bar{\Delta}(|bar{x})$ and $\bar{\psi}_0(\bar{x})$ yield zero.  In demonstrating that each term contributes at $ {\cal{O}} (\mathcal{A}^2 \bar{d}^{0})$ or less, it is often necessary to separately consider the cases where $\bar{\Delta}(\bar{x})$ is localized on a scale of $\bar{w}$ (that is, $\bar{d}^0$) or in which it is spread over a parametrically larger fraction of the unit cell since the reason that the contribution to the energy is ${\cal{O}} (\mathcal{A}^2\bar{d}^{0})$ or less  depends on the scale over which $\bar{\Delta}(\bar{x})$ is spread.  However, while the reason that a given contribution is parametrically small may depend on the scale of $\bar{\Delta}(\bar{x})$, the fact that it is small does not.  

The fact that all contributions at $\mathcal{O}(\delta)$ either cancel or are parametrically of $ {\cal{O}} (\mathcal{A}^{2}\bar{d}^{0})$ or less, shows that the our ans\"{a}tz was self-cosistent to the order at which we are working. Therefore, the energy per unit cell up to parametrically small correction is
\begin{equation}
\frac{\bar{E}}{N_{c}}=\frac{\bar{E}_{0}}{N_{c}}+\delta^{2}\frac{\bar{E}_{2}}{N_{c}}\, ,
\end{equation}
where $\frac{\bar{E}_{0}}{N_{c}}$ is the leading order contribution, which is stated in Eq. (\ref{intenergy}) and $\frac{\bar{E}_{2}}{N_{c}}$ is the energy at $\mathcal{O}(\delta^{2})$.

\section{Comparison with direct numerical solutions}
In this section, we compare the result of the low-density baryon crystal energy to the numerical results based on the  Hamiltonian approach of Ref.~\cite{barak}, which as noted before is valid for large  $N_c$ at any quark mass.  Below, we plot the interaction energies for three different quark masses: $\frac{m_{q}}{\sqrt{\lambda}}=10,40,100$. We plot interaction energy (per color) versus the baryon density $b=\frac{B}{L\sqrt{\lambda}}$, where $B$ is the baryon number and $L$ is the physical size of the box. The points represent the results from numerical finite density calculations based on Ref.~\cite{barak}; the lines represent the results based on our heavy-quark-low-density formalism.

The calculation was done for $M=10$ and $L_{s}=130$ at a fixed volume for each of the quark masses. The $\frac{m_{q}}{\sqrt{\lambda}}=10$ calculation was done in a box of size $L\sqrt{\lambda}=4\sqrt{2\pi}$ and the $\frac{m_{q}}{\sqrt{\lambda}}=40,100$ calculations were performed in a box of size $L\sqrt{\lambda}=2\sqrt{2\pi}$.

Obviously, there are numerical corrections to the interaction energy associated with discretization in space and the fact that $M$ is finite. For the large quark masses involved the corrections associated with choosing a finite $M=10$ and a non-infinite number of lattice sites $L_{s}=130$ are relatively small~\cite{barak}. The discretization correction depends on the number of lattice points that define the wave function of a single baryon. Since, we probed different densities in a box of the same size (for each quark mass), the error associated with discretization is comparable for different baryon densities.

\begin{figure}[ht]
\includegraphics[width=3.2in]{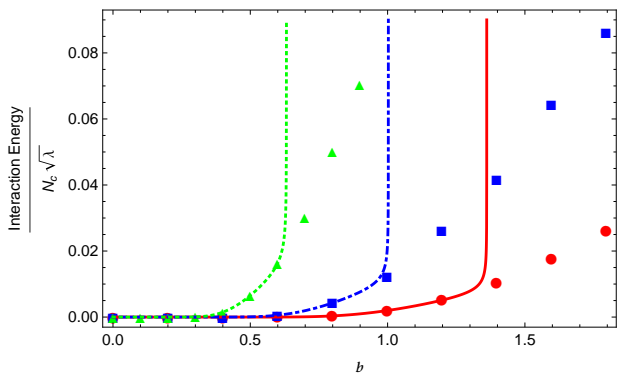}
\caption{Interaction Energy with $\frac{m_{q}}{\sqrt{\lambda}}=10,40,100$ are represented by triangle (dashed), square (dash-dotted) and circle (solid) respectively.  The points are numerical calculations  for the finite baryon density 't Hooft model using the method of Ref.~\cite{barak} and the lines are the result from the heavy-quark-low-density treatment developed in this work.  $b\equiv\frac{B}{L \sqrt{\lambda}}$ (where $B$ is the baryon number in a box, and L its length)  is a dimensionless measure of baryon density.}
\label{interactionfigure}
\end{figure}


We find that the agreement between the heavy-quark, low-density calculation of Eq.~(\ref{intenergy}) and the results based on Ref.~\cite{barak} is remarkabably good for small $b$.  Indeed, it is surprising just how well it works. In deriving Eq. (\ref{intenergy}), rather strong assumptions were made -- including the fact that contributions of $\mathcal{O}(\mathcal{A}^{2}\bar{d}^0)$ were neglected as compared to terms of $\mathcal{O}(\mathcal{A}^{2}\bar{d}^{\frac{1}{4}})$.  One might expect these to become very important as $b$ increases.  Note from Fig.~\ref{interactionfigure} that the approximation breaks down dramatically at certain fixed values, which depends on the quark mass.  As noted above, this break down occurs at $\bar{d} = 2\bar{\epsilon} $.   It is remarkable, however, how accurate the large-mass-low-density expression remains even quite close to the break down value.

\section{Discussion}

In this paper,  the nonzero baryon number sector of large $N_{c}$ QCD in $1+1$ dimension in the limit of large quark masses was explored. The energy and density profile of individual baryons for different quark masses and the interaction energies of low density baryon crystals were computed.   These observables were calculated directly using Witten's mean-field approximation, which simplifies greatly for infinite quark masses~\cite{witten} since the result depends entirely on interactions involving single gluon exchanges. This approach, although valid in a limited regime of quark masses has the virtue of providing a simple intuitive picture for nonzero baryon density which complements the numerical approach Ref.~\cite{barak}.   The computed mean-field energy and baryon density profiles  compare very well with ones calculated numerically based on the Hamiltonian of Ref.~\cite{barak}. The agreement in the large quark mass regime is excellent, and demonstrate the  significance of the effect of all the gluon windings even in cases where the baryon density is exceptionally small at the boxes' edge.

The analysis of low density baryonic matter developed in this paper relied on an {\it ad hoc} assumption that the hidden-color states -- in which the wave function cannot be factorized into the product of color singlet baryons, play a subleading  role at large $N_{c}$.  This assumption is highly plausible; however, unlike for the single baryon case, has not been explicitly demonstrated to be correct \cite{3+1}.  The fact that it agrees so well with the numerical solution, which was derived to be valid at large $N_c$ without making such an {\it ad hoc} assumption, is strong evidence that the assumption is in fact valid. Analogous reasoning was used for the case of large $N_{c}$ QCD with heavy quarks in 3+1 dimensions \cite{3+1}  in the low density regime; the success of the approach in 1+1 lends credence to the 3+1 calculation for which there is no numerical cross-check.

Finally, the success of our low density approximation with its \textit{ad hoc} assumption in matching the numerical results based on Ref.~\cite{barak}, which was designed to work at large $N_c$ without such {\it ad hoc} assumptions, suggests a way forward in formally establishing the validity of the approximation used here and quite conceivably for the 3+1 dimensional case.  Note that the approach of Ref.~\cite{barak} was based on properties of a class of coherent states.  It is highly plausible that a formal treatment based on the coherent states approach of Ref.~\cite{barak} but taken for the heavy quark limit will yield the formalism developed here and thus replace our {\it ad hoc} assumption with something more solid on theoretical grounds.  Moreover, it is plausible that some extension of such a method may work for the 3+1 dimensional case as well.  While the general 3+1 dimensional case has a very complicated structure even at large $N_c$ due to the gluodynamics, in the heavy quark limit the gluons only contribute to produce static color Coulomb interactions and one might imagine an appropriate coherent state formalism applying for the quark fields.   These possibilities will be explored in future work.

\begin{acknowledgements}
P.A. and T.D.C. acknowledge the support of the U.S. Department of Energy through grant number DEFG02-93ER-40762. N.K. acknowleges the financial support of the I.I. Rabi Scholars Program, Columbia University.
\end{acknowledgements}


\begin{thebibliography}{19}
\bibitem{politzer} H.D. Politzer, Phys. Rept. \textbf{14}, 129 (1974).
\bibitem{borici} A. Borici and A. Frommer, "QCD and Numerical Analysis III," (Birkh\"a user, 2005)
\bibitem{hands} S. Hands, Nucl. Phys. Proc. Suppl. \textbf{106}, 142 (2002).
\bibitem{barbour} I.M. Barbour, S.E. Morrison, E.G. Klepsh, J.B. Kogut and M.P. Lombardo, Nucl. Phys. Proc. Suppl. \textbf{60A}, 220 (1998).
\bibitem{alford} M. G. Alford, Nucl. Phys. Proc. Suppl. \textbf{73}, 161 (1999).
\bibitem{stepanov} M. A. Stephanov, PoS LAT2006, 024 (2006) [arXiv:hep-lat/0701002]. 
\bibitem{tHooft1} G. 't Hooft, Nucl. Phys. B \textbf{72}, 461 (1974).
\bibitem{tHooft} G. 't Hooft, Nucl. Phys. B \textbf{75}, 461 (1974).
\bibitem{bars} I. Bars and M. Green, Phys. Rev. D \textbf{17}, 537 (1978).
\bibitem{thies} V. Schon and M. Thies, Phys. Rev. D 62, 096002 (2000); arXiv:hep-th/0003195; arXiv:hep-th/0008175.
\bibitem{narayanan} R. Galvez, A. Hietanen and R. Narayanan, Phys. Lett. B 672, 376 (2009).
\bibitem{EK} T. Eguchi and H. Kawai, Phys. Rev. Lett. 48, 1063 (1982).
\bibitem{barak} B. Bringoltz, Phys. Rev. D \textbf{79}, 105021 (2009); B. Bringoltz, Phys. Rev. D \textbf{79}, 125006 (2009).
\bibitem{negele} L.L. Salcedo, S. Levit and J.W. Negele, Nucl. Phys. B \textbf{361}, 585 (1991).
\bibitem{axialQCD} F. Lenz, H.W.L. Naus and M. Thies, Annals of Physics 233, 317-373 (1994) 
\bibitem{yaffe} L. G. Yaffe, Rev. Mod. Phys. 54, 407-435 (1982).
\bibitem{witten} E. Witten, Nucl. Phys. B \textbf{160}, 57 (1979).
\bibitem{3+1} T.D. Cohen, N. Kumar and K. K. Ndousse, Phys. Rev. C 84, 015204 (2011).
\bibitem{unsal} M. Unsal and L.G. Yaffe, Phys. Rev. D \textbf{78}, 065035 (2008).  
\end{thebibliography}
\end{document}